# The MAGIC of Data Management: Understanding the Value and Activities of Data Management

Roman Lukyanenko[1]

[1] McIntire School of Commerce, University of Virginia, Charlottesville, VA 22903 USA, romanl@virginia.edu

**Abstract**
In an era dominated by information technology, the critical discipline of data management remains undervalued compared to the innovations it enables, such as artificial intelligence and social media. The ambiguity surrounding what constitutes data management and its associated activities complicates efforts to explain its importance and ensure data are collected, stored and used in a way that maximizes value and avoids failures. This paper aims to address these shortcomings by presenting a simple framework for understanding data management, referred to as MAGIC. MAGIC encompasses five key activities: Modeling, Acquisition, Governance, Infrastructuring, and Consumption support tasks. By delineating these components, the MAGIC framework provides a clear, accessible approach to data management that can be used for teaching, research and practice.

## 1. Living in the Age of Magic

In the 1962 science fiction book, "Profiles of the Future: An Inquiry into the Limits of the Possible", Arthur C. Clarke formulated his famous Three Laws, of which the third law is the best-known: "Any sufficiently advanced technology is indistinguishable from magic." Modern world is increasingly magical, driven by the relentless advances in information technology. Yet, the foundations of this magical world remain ill-understood, and sometimes even neglected.

The modern world is digital. Virtually every aspect of human existence is becoming digitalized or depends in some way on information technology and the information systems built based on it. Just in the last three decades alone, the explosive developments in information technology gave rise to revolutionary changes in the way humans live. Consider some examples. The Internet, which became popular in the 1990s allowed electronic commerce and distributed information exchange. Leveraging the Internet, social media burst into existence in the 2000s, dramatically changing the way humans socialize, obtain and disseminate information.

More recently, artificial intelligence (AI), resulting in such marvels as ChatGPT and driverless cars, has been dubbed "the pinnacle of [human] ingenuity" [17]. AI is estimated to contribute $15 trillion to global GDP by 2030 [43], and could bestow world dominance on the country-leader in AI [18].

With the dramatic expansion of IT, unprecedented demands are placed on computing resources, storage, and bandwidth. Responding to this challenge is quantum computing that harnesses the properties of the smallest elements of matter – photons, electrons, ions - for information processing and communication. The powers of quantum particle are so bizarre and unbelievable that are being called "The God Effect" [10]. Some suggest quantum computing "could be a revolution for humanity bigger than fire, bigger than the wheel" [31].

While new information systems continuously emerge, one thing remains constant. It is hence bigger than the entirety of artificial intelligence, quantum computing, social media, online banking, and the Internet. This thing is *digital information* itself. The essence of information technology and the systems built with it is *information* or *data* (used synonymously here). Without digital information neither driverless cars nor YouTube is possible, and quantum computers are but heaps of expensive junk.

Digital data to be useful and even usable, needs to be *managed* in a systematic way. As a result, data management – the approaches and techniques of handling digital information – is the foundation of modern information technology.

Effective data management is requisite for making informed decisions in organizations and by individuals. It can increase productivity, enhance competitiveness, reduce costs and waste. Conversely, poor data management can lead to significant failures and operational disruptions. For example, a data breach experienced by Equifax in 2017 was caused by inadequate data security measures that allowed hackers to access sensitive information of approximately 147 million people. This breach not only damaged Equifax's reputation but also resulted in substantial financial penalties and a loss of consumer trust.

Data management issues plague even the information technology leaders. Thus, after multimillion dollar investments and much hype, MD Andreson Cancer Center scraped collaboration with an AI system Watson provided by IBM [46]. This is despite IBM's ambitious goal to use AI to cure cancer. The demise of the MD Anderson Watson project was significantly influenced by issues of data quality and data integration. The project struggled with inconsistencies and errors in the data, which hindered the effectiveness of the Watson system in providing accurate and safe recommendations for cancer treatment [52].

Astonishingly, in the world dominated by information technology, a key knowledge related to information technology, the science of data management, continues to take the back seat to the inventions that are possible with it. Data and, by extension, data management continue to be treated as a second-class citizen. As one survey of modern artificial intelligence practices concluded: "Everyone wants to [create of new artificial intelligence], not [do] the data work" [48]. Yet AI means nothing without data. AI is just that: a creative use of mathematics and statistics for extracting further patterns from data. No digital data for AI training, means no new information patterns, means no driverless cars or delivery robots.

There is a relative lack of appreciation for data management, relative to the technologies that are made possible with it, such as AI or quantum computing. For example, since the advent of AI tools such as ChatGPT, universities across the globe rushed to develop new courses in AI. By contrast, fewer new courses on data management were created at the same time. Data management for AI, however, is not the same as for, for example, social media or relational databases. Worse still, in some cases with impeccable data management, the use of artificial intelligence may be entirely unneeded. Massive investments in tools, energy and human capital can be avoided if the data were to be stored in a manner that directly permitted the realization of goals set for artificial intelligence.

There are many reasons why data management takes the back seat despite being a requisite for any type of information technology application. One is a misconception that the approaches and techniques of data management are well-established and can be taken for granted. A simple query for the failures related to the improper handling of data can quickly dispel that myth. Furthermore, the methods techniques and tools of data management are continuously evolving. Data management today is no longer a matter of setting up a relational database following well-known data modeling techniques. Many organizations grapple with the challenge of taming large volumes of heterogeneous data, experimenting with such new technologies as data lakes and more recently, data lakehouses. However, if not done right, these efforts may yield marginal returns and even be counterproductive, as expensive new storage infrastructure may gradually descend into a messy, disorganized chaos.

Complicating the efforts to give data management due consideration is a lack of consensus on what constitutes data management, what activities are involved, when data management starts and ends (giving rise to other activities). Stemming from this is the lack of accessible, easy to understand conception of what data management is. Effective data management requires a holistic approach, and this calls for a systematic and comprehensive framework.

Our objective is to provide a simple approach for understanding data management. We call it MAGIC. It includes modeling, acquisition, governance, infrastructuring (a new term that more aptly captures related activities often dubbed "storage" or "curation"), and consumption support activities. MAGIC encompasses the key activities of data management, applicable to data management of any kind. If done right, the result of data management can be truly magical.

Below we explain MAGIC, but first provide a general data management background.

## 2. Information Systems and Information Technology

To understand the nature of data management it is helpful to have a general overview of information systems and information technology that underlies the need for managing data.

Information technology is the knowledge of building and using information systems. Building information systems involves a range of activities from system analysis and design to implementation and maintenance. This process typically includes identifying user requirements, creating system specifications, developing software applications, and integrating hardware components. On the other hand, using information systems focuses on the application of these systems to achieve organizational and personal goals, such as improving productivity, enhancing customer service, and gaining competitive advantages.

Information systems comprise of interacting components that collect, store, manipulate, retrieve, exchange, and use data or information.[2] The components of an information system include mandatory elements such as computer hardware and computer software, as well as optional elements like users and other information systems and devices. Examples of information systems include Youtube.com, Microsoft Word, ChatGPT, or a university admissions system.

Computer hardware forms the physical backbone of an information system, comprising devices such as servers, desktops, laptops, and networking equipment that perform essential computing tasks. Computer software, another crucial component, includes the operating systems, applications, and data management systems that enable the hardware to process and manage data effectively. Users, while optional, interact with the system to perform various tasks, input data, and generate and use the outputs. Additionally, other information systems and devices can be integrated to enhance the system's capabilities, enabling more comprehensive data analysis, improved communication, and streamlined processes.

Information systems can integrate with other objects to form higher-order systems. For example, an airline control system is an information system that is composed of a number of other information systems, such as databases and specialized software, but also includes human air traffic controllers. An integral part of an airline control system is a software system called enterprise resource planning (ERP). An EPR integrates various business processes by collecting and organizing data from different departments, allowing for better coordination and data-driven decision-making.

Regardless of configuration, any information system handles data. Data is the input into a system, it is transformed in some way, and is commonly provided as an output, for either human or computer consumption. Data is representations of any object or event in some physical medium. The basic value of data is in its ability to convey something about the object they represent more efficiently, safely or even making it possible at all to know something about the represented object. Considering the latter, it is only through data (such as words, drawn images), that we can know the contents of somebody else's mind. Often information is distinguished from data as data that is placed into a meaningful context such that what data represents is more or less clear. We choose not to distinguish between data and information because we fail to find cases when data is devoid of meaning, following some scholars and practitioners who treat data and information synonymously [2, 37, 54, 60, 62]. We also use data as singular noun. Data is a plural form of the Latin *datum*, but datum form is rarely used. In most cases singular usage appears to sound more natural in a sentence.

By manipulating data as a proxy for the objects it represents, we are able to understand and act upon the world in a way that would be impossible or difficult to do without data. The very value of information technology is in the ability of information systems to handle data efficiently, at scale, in a way that would be difficult for humans to do. A calculator is a simple example of this efficiency, whereas a driverless car is a more advanced example.

To understand how data is handled by an information system, consider the example of an online retail system. When a customer places an order on an e-commerce website (an information system), the server which runs the website receives various pieces of data as input. This can be the customer's name, address, payment information, and the details of the items being purchased. These data are commonly entered into the system through user input forms. The code on the server then processes these data in multiple ways. It verifies the payment information, checks the availability of the items in the inventory, calculates the total cost including taxes and shipping, and updates the inventory database to reflect the reduced stock. Additionally, the system might analyze the purchase data to provide recommendations for similar products or to update customer profiles with purchase history. The transformed data are then used to generate various outputs. For the customer, the system provides a confirmation message and an email receipt containing the details of the purchase. For the warehouse, the system generates a pick list and packing slip to facilitate the fulfillment of the order. Additionally, for managerial purposes, the system may provide reports on sales trends, inventory levels, and customer purchasing behavior, which can be used for further analysis and decision-making.

[2] Information system is a type of system. What is special about this is the fact that systems have emergent properties – properties that arise from the interaction of their base components [36]. In the case of information systems, these emergent properties (e.g., productivity, ability to draw insights from data), can be significantly more valuable than the components that make up these systems.

As the example of an online retail system demonstrates, data are the key component of an information system, and data does not materialize from thin air. Data need to be first collected, and then stored and used. As a concrete physical substance, modern digital information is a collection of pulses of light and electric charges. These substances reside on hardware devices such as smart phones or laptops. When manipulated in a specific way, these pulses or charges can be represented as binary digits of zero and one (hence, we call such data *digital data*). When further arranged in predefined ways, these pulses and charges can form more complex patterns that can be interpreted by humans and computers as text, images, videos and also as programming instructions for further manipulating these patterns. Artificial intelligence is a particularly powerful technique for extracting useful information patterns that can then be turned into medical diagnoses or instructions for driverless cars.

This very simple exposition of the nature of digital information is enough to appreciate that it is not trivial to go from the pulses of light to tweets about Olympics, presidential elections, or the farewell tour of Pink Floyd. Data management is concerned with ensuring that data or information as an input, an internal component, and output of information systems allows information systems and people that use them to realize their goals. Hence, data management ensures that information systems deliver on their magic.

## 3. The MAGIC of Data Management

The MAGIC framework distills, and at the same time, simplifies the understanding of the nature of data management. It is a framework that grew out of a project of curating data management publications at a top scientific journal *MIS Quarterly* [9].

The MIS Quarterly data management curation project conceptualized data management as the 5Cs of Data Management. It defined the field of data management as an area of study that investigates and develops activities and methods to conceptualize, collect, curate, consume and control data to support insight, analysis, and action. This definition and conceptualizations, along with other relevant literature, provided the basis for MAGIC, as it further elaborated on and refined these concepts. Thie details of the project, along with other seminal definitions of data management are provided in Appendix A.

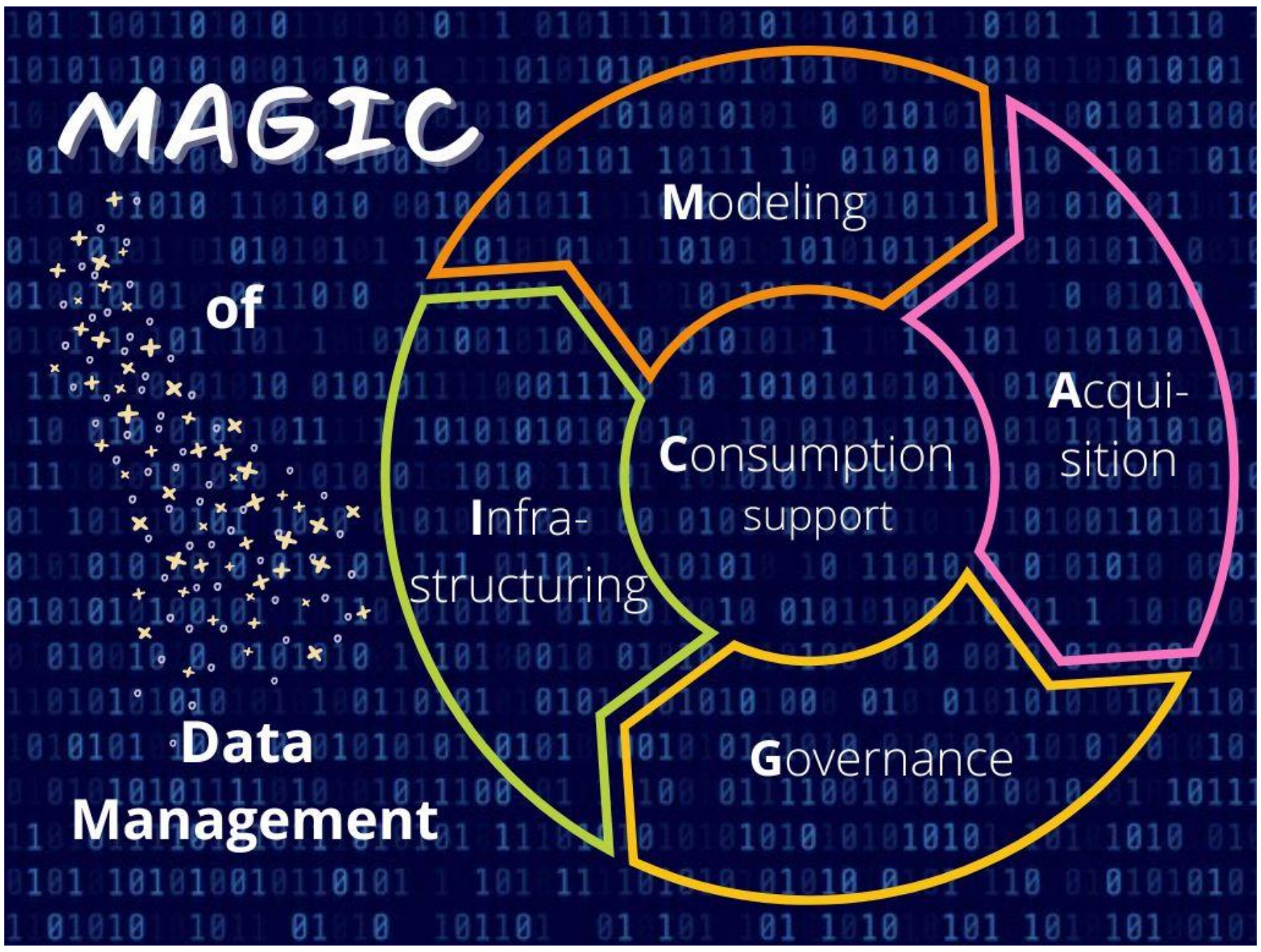

Figure 1. Graphic illustration of the MAGIC framework

The MAGIC framework captures the key activities of data management, involving modeling, acquisition, governance, infrastructuring, consumption support. Figure 1 graphically illustrates the MAGIC framework.

The activities of data management can be performed in the same sequence as the letters of MAGIC, or could be approached in a different order, although modeling invariably is an activity that precedes any other one, whether it is recognized formally or is done at a subconscious level. Furthermore, while these activities emphasize distinct modeling challenges and group related practices, they also partially overlap and feed into one another. For example, formal modeling techniques can be used to design data governance protocols, such as using the Bell-LaPadula computer security model [56]. At the same time, data governance may mandate how formal data modeling is conducted in the organization. Hence, the MAGIC activities must be approached holistically, be in harmony with one another and all must be considered to ensure successful outcomes.

## 3.1. Modeling

Imagine a driverless car. It is a complex information system that includes the vehicle itself, along with the AI-powered software that makes the car drive autonomously. For the AI software to work, it had to be “trained” on millions of examples of human driving, which was used as the original data input. The result of this training is the creation of an AI model, which is digital data that represents the complex rules of driving. Then, as the car begins to drive on the road, it uses its sensors (e.g., cameras, accelerator, lasers), taking in the data generated by the sensors as input, in order to process these data into the instructions for the hardware of the vehicle (accelerate, maintain certain speed, turn at the predetermined moment). This data processing is driven by the AI model built based on the millions of examples of previous driving.

A long, long time before such a car becomes reality, it exists as an idea, as imagination. It exists as a *mental model*. Because we are dealing with not a regular car, but an information system (i.e., the car infused with technology components that process data), we have to include data as an integral component of the driverless car model. Data is as inevitable as wheels and fuel in a driverless car model (see Figure 2 for a mental model visualization).

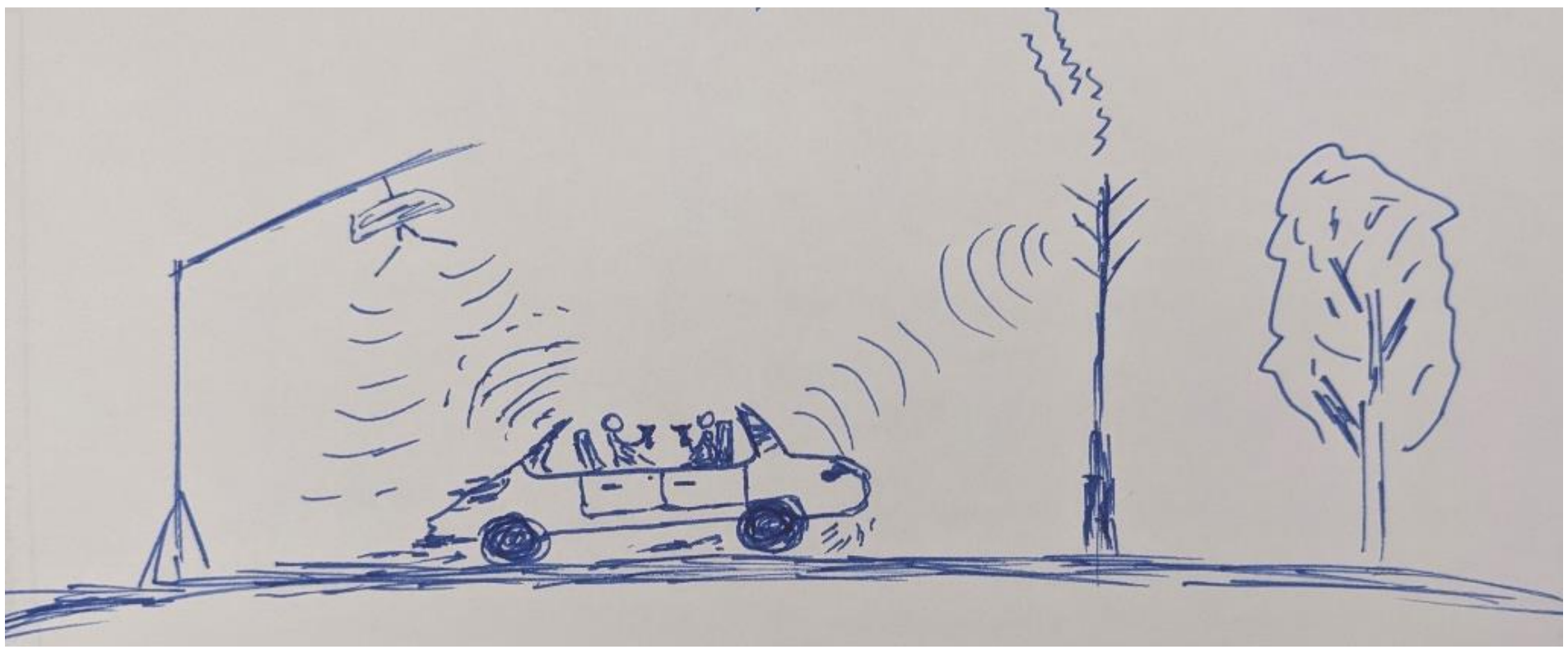
Figure 2. A visualized mental model of a driverless car

Modeling involves making a representation of an object or event for the purposes of understanding, communication, problem-solving and design. A representation is a transformation of some original object into something else. For example, the rule "to stop" can be represented as a red octagon on a post. Representations typically occur in mediums different from their referents (i.e., language can represent physical objects, physical objects can represent abstract mental concepts). This medium shift makes these representations useful (when it is more efficient to work in a medium different than the original), but also creates challenges as perfect translation from one medium to another is impossible. Something gets lost, something else may also be added.[3]

Data modeling, as a particular type of modeling, represents data or any related objects or events for the purposes of understanding, communication, problem-solving and design.

Activities of data management commonly start with modeling and can never escape it. Data modeling can be explicit (when diagrams following strict rules are created), or implicit (when designers think of how the systems should operate, and then follow their ideas when creating programming code and user interface). The very act of thinking about a driverless car involves a mental model, which, if detailed and accurate enough, must account for the indispensable role of data in making such a car a reality.

There are a host of management concerns related to data modeling. These concerns generally seek to ensure that the models provide adequate representations, so that the systems are able to appropriately collect, store and utilize data. These concerns consider that:

- Models are accurate, complete and up-to-date representations so appropriate system components can be built based on them [4, 41, 59];
- Models and modeling are useful and efficient, delivering high returns on the modeling efforts [21, 44, 57];
- Models are accessible, inclusive and easy to use so all relevant stakeholders can participate and shape modeling and subsequent activities [6, 24, 34, 35].

To address these challenge, formal approaches that involve explicit modeling have been developed and continuously evaluated [5]. Some examples of popular modeling techniques include Entity-Relationship Diagrams (ERDs), Unified Modeling Language (UML) and Business Process Model and Notation (BPMN). In addition, new modeling methods and languages are continuously developed. It is important to follow these developments as they may provide better solutions to the modeling challenges.

When building complex or high-stakes systems, explicit models of data are nearly inevitable. These models provide an overview of existing systems, capture the requirements for the new system, describe the kind of data that would be the inputs, elucidate the data transformations and specify the kinds of outputs that the system would generate. These models can then be shown to others for verification, communication, as a roadmap for the creation of programming code and user interface, and they also document the features of the system in a formal way.

---

[3] Recall, data is also a representation. In fact, any physical model is also data. As long as we are able to access mental models in the brains of humans, mental models can also be data. Indeed, models as data frequently become inputs into information systems, such as in model-driven engineering or robotic process automation.

It is especially common to conduct explicit modeling when building databases. Databases store data in a systematic way, which affects the extent to which data is accessible, and secure. To ensure these properties, it is instrumental to follow proven techniques for database modeling which has been in existence since the 1970s [13, 29, 30, 51].

Entity-Relationship Diagrams (ERDs) and Unified Modeling Language (UML) are both powerful tools used in the design and modeling of information systems, though they serve slightly different purposes and are utilized in different contexts. ERDs are primarily used to model the data structure of a system by representing entities (such as tables in a database), their attributes, and the relationships between them [8]. For example, an ERD for a customer management system might include entities like "Customer," "Order," and "Product," with relationships showing how customers place orders and orders include products. ERDs help in visualizing the database schema and are crucial in database design, ensuring that all necessary data relationships are considered before implementation.

On the other hand, UML is a more comprehensive modeling language that encompasses various diagram types to represent different aspects of software systems, including structure, behavior, and interactions [27]. UML includes class diagrams, which are similar to ERDs and used to model the static structure of a system by showing classes, their attributes, methods, and relationships. Additionally, UML features use case diagrams to capture functional requirements, sequence diagrams to model interactions over time, and state diagrams to represent the states and transitions of objects within the system. This versatility makes UML a standard tool for software engineering, providing a unified way to model complex systems from different perspectives, facilitating communication among stakeholders, and supporting the development lifecycle from analysis and design to implementation and maintenance.

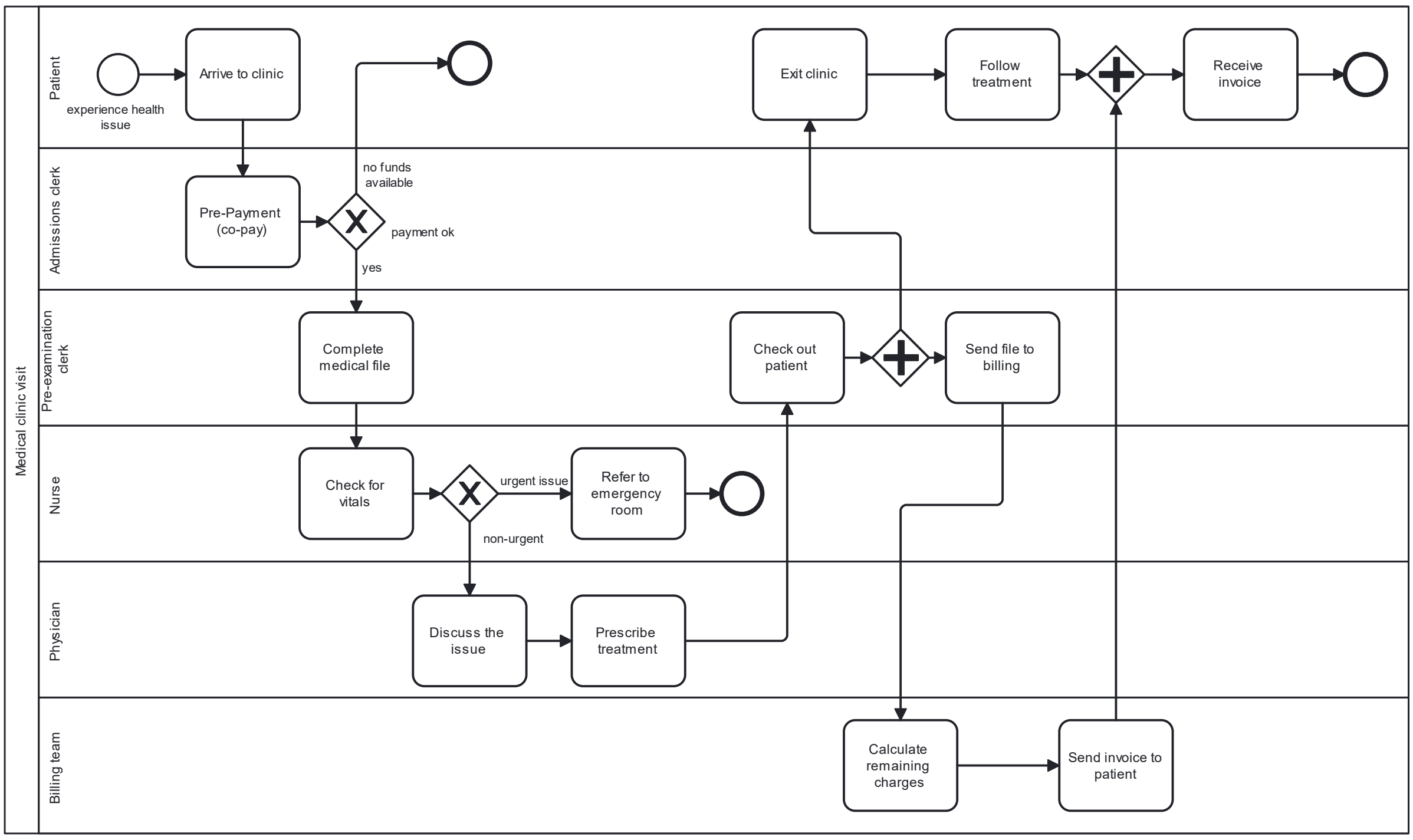

Figure 3. A BPMN diagram of a private clinic visit from the patient's point of view. A simple process reveals surprising twists and turns that may not be apparent without a concrete model.

Business Process Model and Notation (BPMN) is a popular modeling approach for representing business processes that information systems support or implement [12]. It provides a set of symbols that allow stakeholders, including business analysts and technical developers, to visualize and communicate processes clearly and effectively. BPMN diagrams typically include elements such as events, activities, gateways, and flows, as well as data objects, such as databases. These elements enable organizations to explicitly document, analyze, and optimize their business processes. By using BPMN, companies can enhance collaboration, improve process understanding, and drive efficiency in operations. Figure 3 shows a sample BPMN diagram representing a typical private clinic visit process.

Modeling precedes any other activity of data management, as before any other data management activity can be approached, it needs to be first thought of, that is, modeled. This act of thinking can produce an explicit physical model (such as a BPMN diagram dealing with data controls) or remain implicit creating mental models (such as mental blueprint for the kinds of data the system would collect). The MAGIC of data management starts with modeling.

## 3.2. Acquisition

The modeling phase commonly culminates in the implementation of data models into concrete information systems components. This is much like building an actual house after the creation and verification of the construction blueprints. This gives rise to other data management activities: acquisition, governance and infrastructuring.

We begin with data acquisition, as the point when data is born or receives a new life, but it is important to stress that acquisition itself must be subject to appropriate controls. Furthermore, data, to be successfully acquired, requires infrastructure for the data to be stored in. However, acquisition is a separate activity, that, while is conducted under the consideration of the other activities, also shapes them.

Acquisition addresses the issues related to the collection and procurement of data as information systems input. This includes the considerations of the data needed to achieve the desired objectives of the information system, how and where to obtain it, and how to organize the data collection process.

The feasibility of data collection is a major consideration. It is quite possible that there is a mismatch between what is desirable and what is possible with digital data, even in the digital world. For example, it may be desirable to obtain classified documents on an event but be impossible to do. A data management project, therefore, may terminate at the acquisition phase, if it becomes impossible to obtain the desired data economically, timely, safely, ethically or legally.

Assuming it is feasible to collect data, the next question is whether new or existing data should be acquired. This is based on the cost-benefit analysis of collecting data from scratch or obtaining existing data options or the combinations of the two. The information system may collect new data by soliciting data from its users or the environment. For example, a driverless car may obtain its navigation instructions from a smart phone of the user who wishes to get to a particular destination.

Data can also be obtained from another information system, which is made to interact and exchange pre-existing data. Hence, banks providing credit may obtain customer credit reports from credit reporting agencies, such as Equifax or TransUnion. Hospitals and clinics often interact with insurance companies to verify patient coverage and obtain pre-existing medical records. For example, when a patient visits a new healthcare provider, the provider can obtain the patient's medical history, lab results, and previous treatments from other hospitals or clinics using electronic health records systems. E-commerce platforms often interact with payment gateways and shipping services to streamline operations. When a customer makes a purchase, an e-commerce site could with payment processors like PayPal (paypal.com) or AliPay (alipay.com) to validate and process the transaction. Travel websites like Expedia.com or Kayak.com interact with airline reservation systems (such as Amadeus or Sabre) to provide customers with flight options, pricing, and seat availability. Hotel chains use centralized booking systems that interact with online travel agencies like Expedia.com, Booking.com or Airbnb.com to update room availability and manage bookings. In a digital world, there is a high chance that data does not to be collected from scratch but can be obtained from existing sources.

Once the source of data is identified, another data management challenge is ensuring that the right amount of data of sufficient quality is provided as an input for the system so that it can undertake appropriate transformations over these data to realize the set objectives for the systems. Data quality, therefore, is a paramount data collection challenge.

Data quality refers to the condition of data such as its accuracy, completeness, reliability, relevance, and timeliness [61]. Poor data quality can lead to inaccurate analysis, misguided strategies, ruined reputations, and financial losses.

One of the key aspects of data quality is accuracy, which means that the data must represent the real-world entities or events they are meant to depict. Inaccurate data can arise from human error, system glitches, or outdated information. Completeness is another critical factor; it ensures that all necessary data points are present and that there are no missing values that could affect analysis. Additionally, relevance ensures that the data collected is pertinent to the specific objectives and decision-making processes.

Considerations of data quality directly shape the processes for acquiring data and the data acquisition infrastructure. Some guidance for how to shape these processes should come from the prior activity of modeling. In particular, for example, UML diagrams can specify the data fields as attributes, and may also include datatypes and the desired transformations. At the same time, this guidance is incomplete, as many additional design choices need to be made in order to translate this high-level guidance into specific features, such as HTML tags, CSS scripting elements, JavaScript algorithms and so on [38].

Poorly designed user interfaces can significantly contribute to low data quality, by introducing typos and errors due to unclear instructions and labels. When input fields are ambiguously labeled or lack proper guidance, users might misinterpret what data is required. For example, a field simply labeled "Name" without specifying "First Name" or "Last Name" can lead to confusion and incorrect data entry. Similarly, insufficient instructions about the expected format (e.g., date format, phone number format) can result in users entering data incorrectly, leading to errors that could have been easily avoided with better user interface design.

Additionally, the layout and design of the interface can lead to errors or prevent full data from being captured. A cluttered interface with too many elements can overwhelm users, making it difficult for them to locate the correct input fields or buttons, thereby increasing the likelihood of mistakes. Inconsistent placement of elements, such as buttons and fields, can also confuse users, causing them to click the wrong button or enter data in the wrong field. On mobile devices, small touch targets and oversensitive fields can make it hard for users to tap accurately, resulting in typos.

It is advisable to follow best practices in user interface and user interaction design [e.g., 40, 50], not only to ensure that the information systems are easy and enjoyable to use, but also to minimize data quality issues arising from the interface and process design issues. There are many non-obvious solutions that could be powerful in preventing certain types of data quality problems. One is gamified design, that is, designing information systems by following the principles of game design, even if these systems are not games. Such designs can minimize data quality issues by implementing intuitive controls (e.g., colorful and obvious navigation choices) and clear feedback mechanisms (e.g., sounding fanfares when an action is successfully performed).

While data quality is prominent during acquisition, data quality issues permeate the entire data management cycle, and should be considered when modeling, governing, infrastructuring and preparing data for consumption. A common, but often overlooked reason for low data quality is insufficient or biased data modeling, which can ignore certain user requirements, or disenfranchise certain categories of users, making it difficult for them to provide their user input in a faithful manner [33]. Ensuring data quality involves implementing processes and standards that govern data collection, but also storage, and usage which we address later, as part of other activities of MAGIC.

## 3.3. Governance

To ensure data is appropriately acquired, stored and used effectively and responsibly, appropriate controls need to be implemented at different stages of the data life cycle, leading to data governance – a major data management activity. Data governance is generally an organizational activity, but it’s principles can also be applied to personal data management projects, as they deal with legality, ethicality and transparency of data usage.

Data governance involves creating and implementing policies standards, and procedures that ensure data availability, usability, integrity, and security. By implementing robust data governance, organizations and individuals can enhance data quality, comply with regulations, and make better decisions based on reliable data. In organizations, effective data control requires collaboration across various departments, including IT, compliance, and business units, to align data management practices with organizational goals.

One critical aspect of data governance is the definition of data ownership and accountability. Assigning data stewards or owners ensures that there are designated individuals responsible for specific datasets, which helps maintain data quality and consistency. For example, in a healthcare organization, a data steward might be responsible for patient records, ensuring that they are accurate, up-to-date, and accessible while adhering to privacy regulations like HIPAA.[4] This clear delineation of roles helps facilitate data management processes and fosters a culture of accountability within the organization.

Another important component of data governance is the establishment of data policies and standards. These policies dictate how data should be collected, stored, and used, ensuring that all stakeholders adhere to best practices. For instance, a financial institution may have policies in place to govern data encryption, access controls, and data retention periods to comply with regulations such as the GDPR[5] or PCI DSS.[6] By defining clear standards, organizations can reduce the risk of data breaches and enhance the overall security of their data assets.

Security of data infrastructure is a critical aspect of data governance and involves many issues. All sensitive data must be encrypted. Data encryption is a technique for changing the form of data so that it can only be accessed by authorized parties with the correct decryption procedure. An emerging challenge is securing data against the future use of quantum computers. Known as post-quantum cryptography, this is an area of research and practice that designs cryptographic techniques to be secure against the potential threats posed by quantum computers. Quantum computers may solve certain mathematical problems, which classical computers may never be able to solve. This includes problems, such as factoring large integers, the basis for RSA encryption, or computing discrete logarithms, which is used in elliptic curve cryptography. Post-quantum cryptography aims to develop new cryptographic algorithms that remain secure even in the presence of quantum computing capabilities. These algorithms (e.g., lattice-based cryptography, hash-based cryptography, code-based cryptography) are based on mathematical problems believed to be beyond the ability of quantum computers. By preparing early and transitioning to these quantum-resistant algorithms, organizations and individuals can protect their data and communications against future quantum threats, ensuring long-term security in a post-quantum world.

Security also involved the determination of access and permissions over data. Implementing robust access controls, such as multi-factor authentication and role-based access, helps limit who can view or modify the data. Regular security audits and vulnerability assessments identify and mitigate potential weaknesses in the storage system. Additionally, data redundancy and regular backups are crucial for recovery in case of data loss or corruption. Employing advanced threat detection and response tools can also help monitor and address suspicious activities in real-time.

Security also involves protecting the premises, networks and locations of the data centers from unauthorized access and making them robust against psychical attacks, such as arson or theft. It is not uncommon for these issues to be ignored, especially given that the teams working on data management sometimes lack expertise in issues of “physical” security. Holistic data governance involves the consideration of “physical” as well as “digital” aspects of security and ensuring they are tightly choreographed and integrated.

Data quality management is also a key focus within data governance. Organizations need to implement processes for monitoring, assessing, and improving data quality over time. This can involve data profiling, cleansing, and validation techniques to identify and rectify errors in datasets. For example, a retail company might regularly evaluate its customer data to ensure that contact information is accurate and up to date. By prioritizing data quality, organizations can enhance the reliability of their analytics and reporting, leading to better decision-making and strategic planning.

Compliance and risk management are integral parts of data governance as well. Organizations and individuals must ensure that their data practices align with relevant regulations and industry standards. This can involve conducting regular audits and assessments to identify potential compliance risks and implementing corrective actions as needed. For example, a pharmaceutical company must adhere to strict regulatory requirements when managing clinical trial data. By establishing a data governance framework that prioritizes compliance, organizations can mitigate risks and avoid costly penalties.

[4] The United States Health Insurance Portability and Accountability Act of 1996, HIPAA, is a federal law that mandates protecting sensitive health information from being disclosed without the patient's consent or knowledge.

[5] European Union’s General Data Protection Regulation, GDPR, governs how the personal data of individuals in the EU may be processed and transferred.

[6] The Payment Card Industry Data Security Standard, PCI DSS, is a set of guidelines designed to help organizations that handle credit card information to keep that information safe and secure.

Effective data governance ensures regulatory compliance, which is increasingly important in the face of stringent privacy protection laws like the GDPR, the CCPA[7] or PIPL.[8] Companies that implement comprehensive data governance strategies can ensure compliance with these regulations, avoiding hefty fines and legal repercussions.

An emerging consideration is data transparency, which emphasizes openness about data collection, processing, and usage practices. Organizations and interested individuals should clearly communicate their data policies, including how data is collected, stored, shared, and used. Providing users with easy access to their data and the ability to correct or delete it fosters a sense of control and trust. Additionally, data minimization is a key principle that advocates for collecting only the data that is necessary for a specific purpose, thereby reducing the risk of misuse and ensuring compliance with privacy laws.

Finally, effective communication and training are essential components of data governance. Organizations must ensure that all employees understand the importance of data governance and are equipped with the knowledge and skills to adhere to data policies and procedures. This can include conducting training sessions, workshops, and awareness campaigns to promote a data-driven culture. For instance, a technology company might implement a data control training program for its employees to ensure that they understand data handling best practices and the importance of data privacy. By fostering a culture of data governance, organizations can empower their teams to make informed decisions and contribute to the overall success of the organization.

## 3.4. Infrastructuring

Infrastructuring is a new term in data management discourse. It aptly captures an important data management activity in a manner that traditional concepts do not. When data is collected or procured from external sources, data needs a place to persist for a desired period of time. As a result, curation [9] or storage data have been suggested as data management activities. However, curation involves careful selection and organization that some modern storage solutions (e.g., data lakes or random-access memory, discussed later) do not require. Furthermore, considerations of data collection and acquisition infrastructure, the location of data centers, their layout, issues of communication and networks, go beyond the mere considerations of storage or data curation. All these are captured in the activity of infrastructuring – development and implementation of physical and organizational structures and facilities needed to capture and store data.

An existing infrastructure needs to be in place before data can be acquired. When discussing acquisition, we touched on the issue of the design data collection interfaces and processes. Once their shape and form has been decided, they need to be implemented into specific programming code and be stored and managed on the computer hardware (e.g., on premise server or on the cloud). These interfaces deliver data from client computers (laptops, smart phones, Internet of Things sensors) to the host systems (commonly different from the client) that store data temporarily or permanently.

The physical aspects of data storage refer to the components of the infrastructure used to store digital information. This includes various types of hardware, such as random-access memory, hard disk drives (HDDs), solid-state drives (SSDs), tape drives, and cloud storage solutions, each with their own advantages and limitations.

Even if data is discarded without being used permanently, it needs to be held somewhere, in order to be used at the moment of collection. Commonly, such storage involves volatile mechanisms, such as random-access memory. This memory is typically flushed as soon as the data is no longer needed, or the new data replaces the old. Nevertheless, in most information systems, data is stored more permanently. HDDs are typically used for their large storage capacities and cost-effectiveness, making them suitable for bulk data storage. In contrast, SSDs offer faster read and write speeds, making them ideal for applications requiring quick access to data. Tape drives, while slower, are often used for long-term archiving due to their high capacity and longevity. As data storage needs grow, many organizations are increasingly turning to cloud storage, which provides scalable solutions without the need for significant physical infrastructure investments.

[7] The California Consumer Privacy Act of 2018, CCPA, gives consumers more control over the personal information that businesses collect about them and provides guidelines for the organizations on how to implement this law.

[8] The China Personal Information Protection Law of 2021, PIPL, provides rules for the processing of personal and sensitive information including legal basis and disclosure requirements.

To systematically organize the data to be stored for efficient access and secure control, an ecosystem of different technologies has been developed. To store transactional and operational data, many organizations rely on relational databases, a traditional storage technology. At the same time, an alternative family of databases, NoSQL is steadily gaining market share, due to its benefits related to scalability and flexibility. To store data for analytical reporting needs, the traditional solution of a relational data warehouse continues to be popular. At the same time, to store massive volumes of heterogeneous data and improve support for the rapidly exploding field of artificial intelligence, data lakes emerge as a valuable approach.

Relational databases are a traditional and still widely used form of data storage technology that organizes data into tables, or relations, which can be linked based on predefined relationships. Relational database model of storage was invented in 1970s [11] and since then have been integral part of world's data infrastructure, proving to be safe and reliable. These databases utilize Structured Query Language (SQL) for defining and manipulating data, making it easy to perform queries (to read, write, and modify data). Well-known examples of relational database vendors include Microsoft SQL Server, MySQL, PostgreSQL, and Oracle Database. Relational databases are ideal for applications requiring structured data and ACID (Atomicity, Consistency, Isolation, Durability) properties to ensure reliable transaction processing and high data integrity. Their schema-based design ensures consistency but can limit flexibility in handling changing data structures and unstructured or semi-structured data.

Responding to the shortcomings of relational approach in handling large volumes of unstructured or semi-structured data storage, NoSQL databases emerged as a major alternative. These databases offer flexible schemas and are designed for scalability and high performance [3, 22, 23, 47]. NoSQL databases are categorized into four principal types based on their data models, including document stores (e.g., MongoDB), key-value stores (e.g., Redis), column-family stores (e.g., Apache Cassandra), and graph databases (e.g., Neo4j). Unlike relational databases, NoSQL databases do not require a fixed schema, making them suitable for applications with dynamic data requirements, such as social media platforms, real-time analytics, and artificial intelligence. Figure 4 provides an example of a graph data structure.

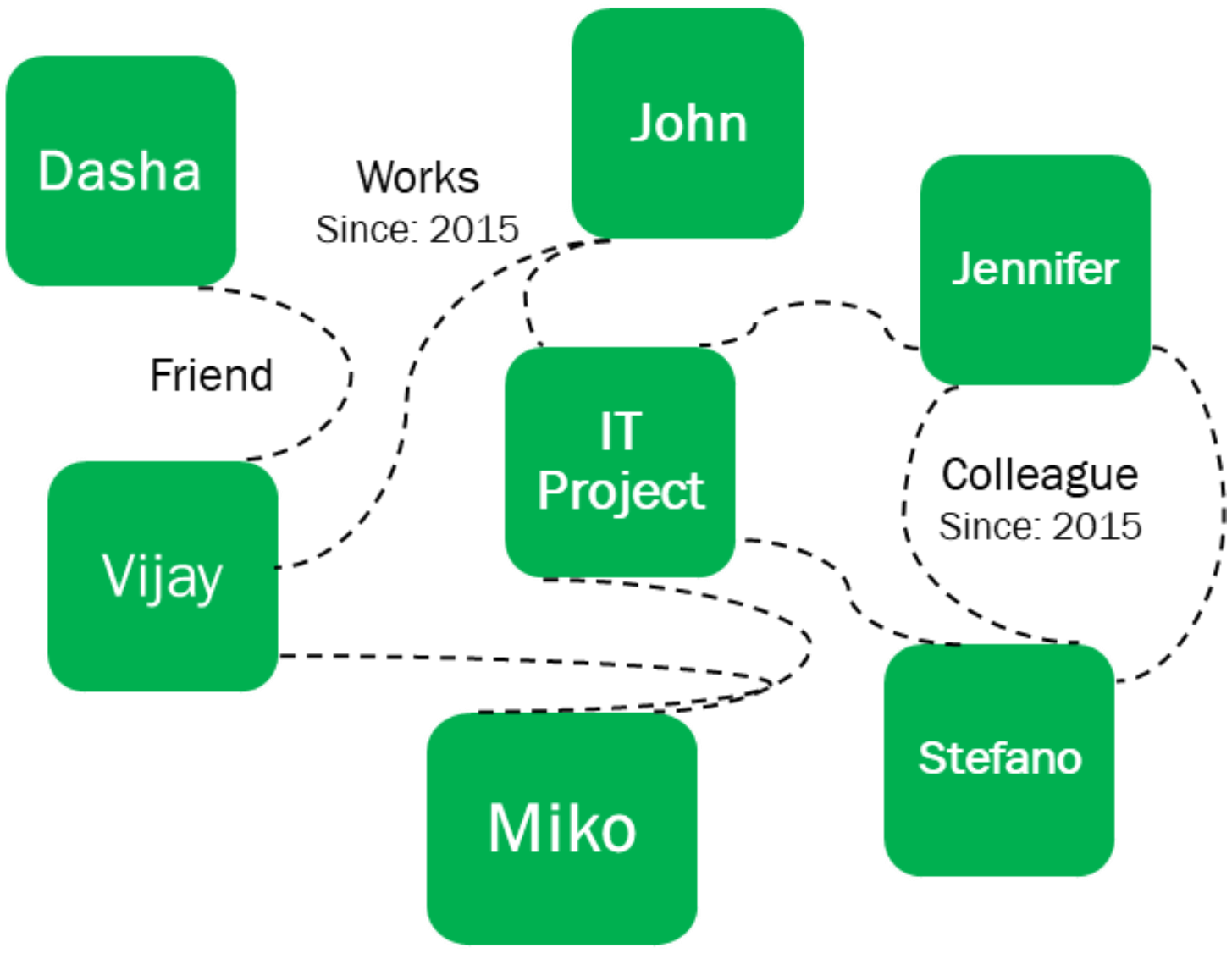

Figure 4. Sample graph model showing nodes (people, project) and edges (relationships). In some graph databases, the edges can have properties.

Data warehouses are specialized storage systems designed for analytical processing and reporting. They consolidate and store large amounts of structured data from various sources, enabling complex queries and business intelligence activities [26]. Data warehouses use Extract, Transform, Load (ETL) processes to ensure data is clean, consistent, and optimized for analysis. They support Online Analytical

Processing (OLAP) for multidimensional analysis and are optimized for read-heavy operations. Examples of data warehouse solutions include Amazon Redshift, Google BigQuery, and Snowflake. By providing a centralized repository of historical data, data warehouses empower organizations to make data-driven decisions and gain insights into business trends and performance.

Data warehouses, especially of the traditional types, operate on fixed schema and struggle to support heterogeneous data, such as text, images and videos. These limitations gave rise to data lakes and related technologies. Data lakes are designed to store vast amounts of raw, unstructured, semi-structured and structured data in its native format [19, 20, 53]. Data lakes support a schema-on-read approach, allowing data to be interpreted and structured at the time of analysis, rather than during ingestion. This flexibility makes data lakes ideal for big data analytics, machine learning, and data science projects. Technologies such as Apache Hadoop and Amazon S3 are commonly used to build data lakes. Popular data lake platforms include Amazon Lake Formation, Microsoft Azure Data Lake Storage, Google BigLake, Cloudera Data Platform, Databricks Delta Lake and Dremio. By providing a unified storage repository, data lakes enable organizations to perform advanced analytics and uncover insights from diverse datasets, but they require robust data governance and management practices to ensure data quality and usability.

A major concern is the potential for data lakes to become data swamps. Given that data lakes can store vast volumes of sometimes loosely connected and ill-organized unrelated data, if not carefully managed, they can quickly become cluttered and messy, making it difficult to find the relevant data and connect it with other data. Consequently, an emerging practice is to pair a data lake with a data warehouse, which can impose structure upon the data lake data objects. This approach is called data lakehouse [25]. Databricks is a prominent data lakehouse vendor. Data Lakehouse is the most recent episode (with relational versus NoSQL being another major episode) in the perennial struggle between structure and no structure in data storage.

The physical layout of data centers plays a crucial role in data storage. Data centers must be designed to optimize performance, reliability, and security. This includes considerations such as temperature control, power supply, and redundancy to prevent data loss in case of hardware failure. Furthermore, physical security measures, such as access controls, surveillance, and fire suppression systems, are vital to protect sensitive data from unauthorized access or damage. As the volume of data continues to increase, organizations must carefully evaluate their physical storage solutions to ensure they can efficiently store, manage, and protect their data while accommodating future growth.

Networking and connectivity are part of data management, insofar as they shape and constrain the ability to exchange information. An important connectivity infrastructure is application programming interfaces (APIs). APIs are protocols that enable information systems to communicate with each other to exchange data. With the diversification of the information storage landscape, increasingly data access and manipulation are handled via APIs (as opposed to the traditional SQL). Consequently, the design of APIs can become either a facilitator or a bottleneck in data exchange [53].

A key emerging consideration when storing data is the impact on the environment. Green data storage refers to environmentally friendly practices and technologies used to manage data storage in a way that minimizes energy consumption, reduces carbon footprints, and promotes sustainability [45]. As the demand for data storage continues to grow exponentially, traditional data centers have come under scrutiny for their significant energy usage and environmental impact. Green data storage aims to address these challenges by adopting strategies that not only enhance efficiency but also reduce the ecological footprint of data storage operations.

One of the primary components of green data storage is the use of energy-efficient hardware and infrastructure. This includes employing SSDs over traditional hard disk drives, as SSDs typically consume less power and generate less heat. Additionally, data centers are increasingly utilizing energy-efficient cooling systems and renewable energy sources, such as solar or wind power, to further reduce their energy consumption. Virtualization and cloud storage solutions also play a crucial role in green data storage by optimizing resource utilization, allowing multiple virtual servers to run on a single physical server, thus reducing the need for additional hardware.

Infrastructuring involves the implementation of data governance plans and procedures related to security. This involves putting in the physical infrastructure and writing required programming code to ensure data security, access and permissions, and installation and maintenance of physical security elements, such as buildings, locks, implementation of physical security controls.

Infrastructuring is an ongoing process, as it involves frequent modifications to the data facilities in order to accommodate the expanding data usage demands and evolving requirements. Hence, it is not uncommon to spawn a data lake into a data lakehouse [25], when it becomes evident that the original data lake was descending into a data swamp (issues we discussed later). Changes to the infrastructure are also common in an attempt to optimize the performance of information retrieval or to minimize the environmental impact. Hence, while intimately connected with acquisition, it is an activity that also stands on its own.

## 3.5. Consumption support activities

Just because data is stored and organized, does not mean it can be accessed and used. The final activity of data management seeks to make data usable, supporting its eventual consumption.

There are many different approaches to enhance data for decision-making, which can be implemented as part of data management efforts. What separates them from data manipulation activities conducted as part of data consumption is their oddness to city toward a particular task. These are efforts to ensure data can support organizational or personal decision-making broadly, in a variety of cases and for a variety of purposes. Metaphorically, these activities can be thought of as an ocean tide that raises all the boats at the same time. By conducting these activities future data manipulation and consumption becomes more effective and efficient. Here, we highlight some of these activities.

**Query optimization.** A traditional approach to retrieving stored information from databases is with the use of queries written in data access languages. The most common data access and manipulation language is SQL (Structured Query Language), created originally to support relational databases. With the rise of NoSQL databases, new data access languages emerged, such as Dynamo Query Language (DQL) for DynamoDB, MongoDB Query Language (MQL), and Cypher for Neo4J. The most developed optimization approaches, however, continue to be for SQL.

Query optimization is the process of improving the performance of a database query by minimizing resource usage and execution time while maximizing efficiency. This involves analyzing and transforming the SQL statements that are executed against a database to ensure that they retrieve the desired results in the most effective manner. Query optimization can include techniques such as rewriting queries, creating or modifying indexes, and adjusting database configurations to enhance the execution plan that the database engine follows when processing the query.

An essential aspect of query optimization is the use of execution plans, which outline the steps the database will take to execute a query. The database management system (DBMS) generates these plans based on various factors, including the structure of the data, existing indexes, and the complexity of the query. Optimizers evaluate different strategies to determine the most efficient way to retrieve the requested data, often considering the cost of operations like joins, scans, and filters. By employing effective query optimization techniques, organizations can significantly reduce response times for complex queries, improve overall system performance, and enhance user experience, ultimately enabling better data-driven decision-making.

**Data quality assessment and improvement.** Assessment and optimizing data for future uses involves continuous assessment of the quality of stored data, including its accuracy, completeness, consistency, relevance, and timeliness. A common technique is data profiling, which involves analyzing the dataset's structure, types, and content to identify potential issues. This can include checking for missing values, duplicate records, and outliers that may indicate data entry errors or anomalies. Tools and techniques such as descriptive statistics and data visualization can be employed to gain insights into the dataset and highlight areas that may require attention.

Ideally, organizations should establish specific criteria and benchmarks for assessing data quality based on the intended use of the dataset [32]. This involves comparing the dataset against authoritative sources or standards to ensure accuracy and consistency. For instance, if a dataset is used for demographic analysis, it should align with recognized demographic classifications. Additionally, relevance should be assessed by evaluating whether the data meets the current needs of the organization or project. Finally, the timeliness of the data must be considered, ensuring that it is up-to-date and reflects the most recent information available. By systematically evaluating these dimensions, organizations can identify data quality issues, prioritize remediation efforts, and ultimately enhance the reliability and usability of their datasets for decision-making.

Part of data quality review and assessment is the determination of whether data needs to be archived or deleted, if no longer needed. Archived data can be stored on slower-to-read mediums and be additionally compressed to further optimize storage efficiency and decrease the overall storage and environmental footprint.

**Data integration and interoperability.** Data integration and interoperability emerges as an essential data management challenge because neither data nor the systems that handle data exist in isolation. To be usable, modeled, acquired, and housed data needs to be made available to its users. This gives rise to data integration – integration of data with data users and other systems - as a common data management activity.

Despite its importance and ubiquity, data integration is often overlooked as a data management activity. Indeed, none of the definitions of data management shown in Appendix A mention it explicitly. Yet, insufficient attention to data integration issues frequently results in catastrophic failures. Among the issues that doomed the MD Andreson-Watson project (mentioned earlier), was failure to tame the complexity of integrating vast amounts of medical records from diverse sources. In a similar vein, generative AI tools such as ChatGPT hallucinate, that is makeup nonsensical information [28]. AI hallucination is often due to the failure to appropriately integrate different data sources used for training these technologies.

To make data useful integration encompasses version control, metadata management, and data lineage tracking. Version control ensures that changes to data models are documented and can be reverted, if necessary, while metadata management provides context and information about the data, such as its source, structure, and purpose. Data lineage tracking helps understand the flow of data throughout its lifecycle, enabling to trace back to the origins of the data and ensuring compliance with data governance policies.

Metadata management involves the systematic organization and maintenance of metadata, which is data that describes and provides context for other data. This process is crucial for enhancing data governance, quality, and usability across an organization. Effective metadata management helps users understand the origins, meanings, and relationships of data, enabling them to make informed decisions and utilize data more effectively. For example, in a healthcare organization, metadata might include information about patient records, such as the data's source, the structure of the data fields, and how often the data is updated. This contextual information ensures that healthcare professionals can trust and utilize the data for accurate patient care.

A key component of metadata management is the creation of a centralized metadata repository or catalog. This repository acts as a single source of truth for all metadata within an organization, making it easier for stakeholders to find and access relevant data. For instance, a financial services firm might implement a metadata management system that catalogs various datasets, including transaction records, customer profiles, and regulatory compliance data. This system would provide detailed descriptions, data lineage, data classification, and data ownership information, enabling analysts and compliance officers to quickly locate the necessary data and understand its context and constraints. By centralizing metadata, organizations can improve collaboration among teams, reduce redundancy, and enhance data discovery.

Data integration can be facilitated by the use of domain ontologies. Ontologies are formal representations of knowledge within a specific domain, defining the concepts, relationships, and rules that govern the data in that domain [21, 39]. They provide a shared vocabulary and a structured framework for understanding and integrating diverse datasets, facilitating interoperability among various systems and applications. By providing a common understanding of the data, ontologies help to bridge gaps between disparate systems, making it easier to combine and analyze data effectively.

One advantage of using ontologies for data integration is their ability to model complex relationships. For example, in the healthcare sector, an ontology might define relationships between patients, treatments, medications, and outcomes. By mapping data from various sources—such as electronic health records, clinical trial databases, and insurance claims—onto this ontology, healthcare providers can achieve a unified view of patient information. This integration allows for more comprehensive analyses, improved decision-making, and better patient outcomes.

Ontologies also support data interoperability by enabling semantic reasoning, which allows systems to infer new knowledge based on existing relationships defined in the ontology. For instance, if an ontology specifies that "a cardiologist is a type of physician" and "a physician can prescribe medication," a system using this ontology could automatically infer that "a cardiologist can prescribe medication." This capability enhances the ability of different systems to work together and share insights.

**Data enrichment.** The final activity which is becoming increasingly popular is data enrichment, the transformations and additions to the original data set which generally increase its ability to support decision-making, analysis and actions. For example, an organization that relies extensively on machine learning AI may engage in generic feature transformation and engineering efforts.

Feature engineering involves the creation and transformation of input data into meaningful features that can improve the performance of machine learning models. For example, in a dataset containing timestamps of transactions, one might extract features such as the hour of the day, day of the week, or even holidays to capture temporal patterns. In a dataset with text data, techniques like extracting word counts, n-grams, or sentiment scores can turn unstructured text into structured features [16]. When values are missing in a data set, these values may be imputed [7], or efforts may be undertaken to obtain the missing values.

In order to make data AI-friendly for a variety of tasks, feature engineering may be conducted before data is used by data scientists and AI engineering engineers. By enriching the original dataset with relevant features, feature engineering can significantly boost the accuracy and interpretability of machine learning models, leading to better decision-making and insights. Naturally, these efforts need to be considered in the context of other priorities, such as the cost and energy impact of storing additional data. At the same time, generic feature engineering may also result in the overall reduction of storage and computational resources, as this is a one-time effort which replaces multiple feature transformation activities by different machine learning teams.

## 3.6. Beyond data management

The outcome of data management is the provision of data which can be used as input into decision-making, analysis and action. Effective, efficient and responsible data management allows organizations and people to unlock the powers of information technology and use it to its full potential. This can propel organizations into significant prominence and can even change the nature of industries and markets. For instance, Netflix's pioneering approach to entertainment spawned a new industry, dislodging the traditional CD and DVD rental giants, such as Blockbuster. This feat was made possible through innovative and robust data management practices that enabled Netflix to share content directly and offer personalized recommendations to millions of viewers. In a similar vein, companies worldwide compete on data [15], gaining competitive edge through better data modeling, acquisition, governance, infrastructuring, and consumption readiness and support.

Data management can significantly enhance operational efficiency. Properly managed data facilitates integration and access across various departments, fostering collaboration and innovation. Amazon's use of data management to streamline its supply chain operations is a prime example. By integrating data from various sources and applying advanced analytics, Amazon can optimize inventory management, reduce delivery times, and improve customer satisfaction.

Data management is not only a business or organizational activity. As digital information shapes the daily lives of virtually everyone on the planet, good data management skills can greatly enhance personal financial and social outcomes. For instance, well-organized folders on a personal laptop for personal documents like bills, receipts, and warranties allow one to quickly find and reference important information and avoid costly bills or tax filing errors. Efficient data management can also reduce information overload by filtering and prioritizing relevant data, allowing people to focus on high-priority tasks and make informed decisions swiftly. Similarly, by simply organizing photos and videos on a smart phone in a meaningful and intuitive way, it becomes easier to share special moments with others. These improvements lead to better time management, enhanced decision-making, and increased overall productivity in daily activities.

Data management does not end when the data is ready for subsequent use. Monitoring and evaluating the effectiveness of data, and, by extension, of data management, fosters learning and improves practices. Hence, data management practitioners should proactively seek feedback on the extent to which data they provided helped to generate personal and organizational value and make any necessary adjustments to the data management practices.

# 4. Conclusions and the Future of MAGIC

In an era increasingly shaped by information technology, the essential discipline of data management often struggles to gain the recognition it deserves, particularly when compared to the groundbreaking innovations it facilitates, such as artificial intelligence, social media, and data analytics. Despite its critical role in enabling these advancements, data management is frequently perceived as a secondary concern, leading to a lack of investment and attention in this vital area. Compounding this issue is the ambiguity surrounding the definition of data management and the various activities it encompasses. This confusion hampers efforts to communicate the significance of effective data management practices to stakeholders across organizations and in society.

To address these challenges, this paper introduces a comprehensive and straightforward framework for understanding data management called MAGIC. This framework comprises five key activities: Modeling, Acquisition, Governance, Infrastructure, and Consumption support. Each of these components plays a crucial role in the overall management of data, providing a structured approach that facilitates clarity and consistency in data-related practices. By delineating these activities, the MAGIC framework not only enhances the understanding of data management but also serves as a valuable tool for teaching, research, and practical application in various organizational contexts.

By employing the MAGIC framework, organizations can foster a more profound appreciation for data management and its pivotal role in harnessing the full potential of information technologies. Furthermore, this framework equips data professionals, researchers, and educators with a common language and methodology for discussing and implementing data management practices. In doing so, we aim to bridge the gap between data management and the technologies it supports, ultimately promoting a more holistic approach to data as a vital asset. Through this initiative, we hope to elevate the conversation around data management and encourage a renewed focus on its importance in an increasingly data-driven world.

# Appendix A: Foundations of the MAGIC framework

The MAGIC framework grew out of a project of curating data management publications at a scientific journal MIS Quarterly [9], performed by a team of data management scholars from across the globe. The project was solicited by MIS Quarterly, a premier peer-reviewed academic journal that publishes research on the management of information systems and is widely regarded as the top journal of the information system discipline.

The MIS Quarterly data management curation project conceptualized data management as the 5Cs of Data Management. It defined the field of data management as an area of study that investigates and develops activities and methods to conceptualize, collect, curate, consume and control data to support insight, analysis, and action. This definition and conceptualizations provided the basis for MAGIC, as it further elaborated on and refined these concepts. First, the MAGIC framework expands conceptualization into the broader activity of modeling. Second, it expands the activity of collecting data to a broader one of data acquisition that also involves procuring existing data from other systems. The somewhat narrowly scoped activity of data curation is replaced with a new concept of data infrastructuring.

The MAGIC framework does not include the activity of consuming data. We believe this is beyond the scope of data management. Data management, as with any area of practice, needs to have well defined boundaries. Data consumption happens as a result and as a consequence of data management, but itself is a separate activity, which involves using the data to drive decisions, insights and actions. However, beyond dispute is the need to support consumption by various data management techniques, such as metadata management. Consequently, MAGIC refined the consumption activity calling it consumption support.

Finally, the activity of control is retained, but its scope is substantially broadened to better account for the evolving data governance practices. Furthermore, a more traditional label of governance is used in MAGIC, in line with the industry and research practices [1].

In addition to the 5Cs of Data Management framework, MAGIC considered leading definitions of data management. Some are provided in Table A1. We considered these definitions to ensure the major facets of data management are covered by MAGIC as understood by the leading thinkers in the field of data management. Comparing the definitions in Table A1 with MAGIC shows that the magic framework encompasses the considerations of the key definitions of data management, whereas any given definition in the table is not as comprehensive as the MAGIC framework.

**Table A1. Popular definitions of data management**

| Definition of data management | Source | Our Analysis |
|---|---|---|
| Data management is the development, execution, and supervision of plans, policies, programs, and practices that deliver, control, protect, and enhance the value of data and information assets throughout their lifecycle | DAMA International [14] | • Focus on the value of data, not data itself<br>• Unclear what activities that relate to data itself are part of data management<br>• Modeling, an essential activity, is not explicitly mentioned (implicit, if the concept of "planning" were to be interpreted as modeling) |
| Data management is the practice of collecting, keeping, and using data securely, efficiently, and cost-effectively. | Oracle [42] | • Using data is generally beyond the scope of data management<br>• No mention of the value or goals of data or reason for data management<br>• Modeling, an essential activity, is not explicitly mentioned |
| Data management is the IT discipline focused on ingesting, preparing, organizing, processing, storing, maintaining, and securing data throughout the enterprise. | Datamation [58] | • Organizational focus, yet data management is also a concern for individuals<br>• No mention of the value or goals of data or reason for data management<br>• Ingestion is one of the many types of data acquisition (along with collection from scratch)<br>• Modeling, an essential activity, is not explicitly mentioned<br>• Organizing is suggested after collection, whereas, in fact, organizing (as part of data modeling) shapes collection |
| Data management is the practice of collecting, organizing, protecting, and storing an organization's data so it can be analyzed for business decisions. | Tableau by Salesforce [55] | • Organizational and business focus, yet data management is also a concern for individuals<br>• Purpose of data management – to make decisions – is explicit<br>• Modeling, an essential activity, is not explicitly mentioned<br>• Organizing is suggested after collection, whereas, in fact, organizing (as part of data modeling) shapes collection |
| Data management is the practice of collecting, organizing, managing, and accessing data to support productivity, efficiency, and decision-making. | SAP [49] | • Organizing is suggested after collection, whereas, in fact, organizing (as part of data modeling) shapes collection<br>• Managing as an activity is circular to the main concept<br>• Focus on the outcomes: productivity, efficiency, and decision-making<br>• Accessing data is an act beyond data management; however, preparing data for efficient access is |